\documentclass[twocolumn,pra,showpacs]{revtex4}

\usepackage{epsfig}
\usepackage{graphics}
\usepackage{amssymb,amsmath,amsbsy}
\usepackage{graphicx}
\usepackage[latin2]{inputenc}
\usepackage{bm}

\newcommand{\e}{\mathrm{e}}                  
\newcommand{\ket}[1]{|#1\rangle}             
\newcommand{\bra}[1]{\langle #1|}            
\newcommand{\braket}[2]{\langle #1|#2\rangle}
\newcommand{\mt}[1]{\mathrm{#1}}             

\newcommand{\wh}{\widehat}

\begin{document}

\title{Ramsey interferometry with oppositely detuned fields}

\author{D.\ Seidel}
\email{dirk_seidel@ehu.es}
\author{J.\ G.\ Muga}
\email{jg.muga@ehu.es}
\affiliation{Departamento de Qu\'{\i}mica-F\'{\i}sica, Universidad del
Pa\'{\i}s Vasco, Apartado Postal 644, 48080 Bilbao, Spain}

\begin{abstract}
We report a narrowing of the interference pattern obtained in an atomic Ramsey interferometer if the two separated fields have different frequency and their phase difference is controlled. The width of the Ramsey fringes depends inversely on the
free flight time of ground state atoms before entering the first field region in addition to the time between the fields. The effect is stable also for atomic wavepackets with initial position and momentum distributions and for realistic mode functions.
\end{abstract}

\maketitle

\section{Introduction}

Ramsey's method of atom interferometry with separated oscillating fields \cite{Ramsey} provides the basis of present primary time standards, as cesium-beam or cesium-fountain setups \cite{cesium_fountain}. Basically, one aims to lock an oscillator exactly to a given atomic transition frequency to achieve stability and accuracy of the oscillator and thus of the clock. The physical quantity that indicates possible deviations from the reference transition as a function of detuning is the excitation probability of the ground state atom after the interaction with two separated field pulses, and this function shows the well-known Ramsey fringes. 
Note that as long as quantum reflections of the atom at the fields can be neglected, the operation of the interferometer in time domain (temporally separated pulses and fixed atom) or in space domain (spatially separated fields and moving atom) is equivalent for atoms moving along classical trajectories, as we shall exploit hereafter.

A central requirement for frequency standards is a narrow interference pattern with respect to detuning, to allow for a precise lock of the oscillating fields  to the atomic clock transition. It has been shown by Ramsey \cite{Ramsey} that the width of the central peak of the pattern is inversely proportional to the intermediate time between the two pulses $T$, in contrast to single-field methods where it is inversely proportional to the field-crossing time. In the case of cesium-beam standards, for example, this implicates the desire of using tall ``fountain'' configurations, which of course have a limit in practice because of space constraints, magnetic field control, temperature homogeneity and other technical aspects \cite{WyWey-Metr-2005}; or very slow (ultracold) atoms in reduced gravity \cite{ultracold}. 
Even though the dependence on the free flight time is far better than on field-crossing time because of the difficulty to implement a homogeneous and stable
field, the free flight occurs for atomic states with an excited component, which should be as stable as possible against radiative decay \cite{decay}.
 
In this paper we report an effect that leads to a considerable narrowing of the Ramsey interference fringes which is stable with respect to the atomic velocity
or field mode functions,
and relies on the free flight of ground state atoms before entering the first field region.  
This effect is based on the use of two field pulses with {\it different} detuning. 
For simplicity of the presentation, we neglect in our calculations the transverse momentum transfer on the atom, which is reasonable for moving atoms and microwave frequencies  or for trapped ions or atoms in the Lamb-Dicke regime and optical fields \cite{BeItWi-PRA-1987,IdKa-PRL-2003}. For a detailed study of these recoil effects in connection with Ramsey interferometry we refer to Refs.~\cite{Borde1}.
For freely moving atoms interacting with optical fields, one is led normally to consider the multi-beam schemes proposed by Kasevich and Chu or by Bord\'e to compensate for the wavepacket separation due to recoil effects, see \cite{Berman-book} for reviews.
However, to suppress transversal momentum transfer one may think on moving atoms in  a narrow waveguide, interacting with optical fields \cite{SeiMu-preprint-2006}. In fact, for a waveguide width of $100\,\mt{nm}$ and for cesium, the energy gap to the first transversely excited state is $\delta E = 2\pi\hbar\times 0.113\,\mt{MHz}$. Now, a minor modification of ref.~\cite{Ion-preprint} to incorporate detuning shows that excitation mainly occurs at the ``Rabi resonances" $\hbar(\Omega^2 + \Delta^2)^{1/2} = \delta E$, where $\Omega$ is the Rabi frequency and $\Delta = \omega_\mt{L}-\omega_{21}$ denotes the detuning between laser frequency and atomic transition frequency. For Rabi frequencies of the order of $2\pi\times 0.016\,\mt{MHz}$ one therefore would have a detuning range $\Delta \approx 2\pi\times (-0.11 \dots 0.11)\,\mt{MHz}$ for which transversal excitation can be neglected. 

\section{Hamiltonian and interaction pictures}

\begin{figure}
\centering
\epsfxsize=8cm
\epsfbox{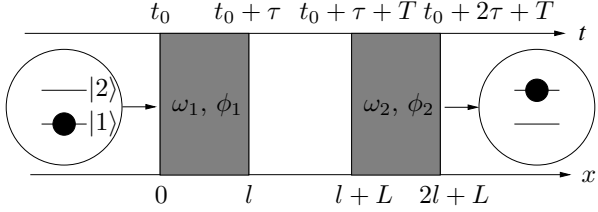}
\caption{Scheme of the Ramsey atom interferometer. The oscillating fields can be either separated in space, interacting with an atom beam (lower axis), or in time, interacting with a trapped ion or atom (upper axis).}
\label{fig:setup}
\end{figure}

We consider the basic Ramsey setup where either a two-level atom in the ground state moves along the $x$ axis and crosses two separated oscillating fields localized between $0$ and $l$ and between $l+L$ and $2l+L$ (see Fig.~\ref{fig:setup}), or the atom is trapped and the two interaction pulses are separated in time. 
In contrast to the standard setting, we allow for different detuning of the two fields with respect to the atomic transition frequency $\omega_{21}$, respectively. The measured quantity is the transmission probability of excited atoms, $P_{12}$, as a function of the detuning $\Delta_j = \omega_j - \omega_{21}$, $j=1,2$, where $\omega_j$ is the frequency of the $j$th field. 
The Hamiltonian describing the moving atom reads in the dipole and rotating-wave approximation and in the Schr\"odinger picture
\begin{multline}
H(t) = \frac{\wh{p}^2}{2m} + \hbar\omega_{21} \ket{2}\bra{2}\\ + \sum_{j=1,2} \frac{\hbar\Omega_j(\wh{x})}{2} \left(\ket{1}\bra{2} \e^{i\omega_j t} + \ket{2}\bra{1} \e^{-i\omega_j t}\right),
\end{multline}
where $\Omega_j(x)$, $j=1,2$, are the Rabi frequencies of the two spatially localized fields and the hat $\widehat{~}$ is used to distinguish operators from the corresponding c-numbers. It is important in the following that the time origin has been chosen in a way that the two fields are in phase at $t=0$. For trapped ions or atoms, the kinetic energy term would be absent and the Rabi frequencies would be time-dependent, which will not alter the results within the following treatment.
The crucial point to note is that for $\omega_1 \neq \omega_2$ there is no interaction picture for which this time-dependent Hamiltonian can be made time-independent, as it is the case for $\omega_1 = \omega_2$. Thus, the quantum mechanical probabilities, and in particular $P_{12}$, will also be time-dependent.

In the atom-adapted interaction picture with $H_0 = \hbar\omega_{21}\ket{2}\bra{2}$ one has 
\begin{equation} \label{eq:HI1}
H_\mt{I1}(t) = \frac{\wh{p}^2}{2m} + \sum_{j=1,2} \frac{\hbar\Omega_j(\wh{x})}{2} \left(\ket{1}\bra{2} \e^{i\Delta_j t} + \ket{2}\bra{1} \e^{-i\Delta_j t}\right).
\end{equation}
In this interaction picture, the atom-field interaction is zero in between the fields and therefore we will favor it for the calculation of $P_{12}$.
\paragraph*{Remark:} One may also choose a field-adapted interaction picture with respect to the first or the second field by setting $H_0 = \hbar\omega_j \ket{2}\bra{2}$, which leads for $j=1$ to
\begin{multline}
H_\mt{I2}(t) = \frac{\wh{p}^2}{2m} - \hbar\Delta_1 \ket{2}\bra{2} + \frac{\hbar\Omega_1(\wh{x})}{2} \left(\ket{1}\bra{2} + \ket{2}\bra{1}\right)\\ +  \frac{\hbar\Omega_2(\wh{x})}{2} \left(\ket{1}\bra{2} \e^{-i(\Delta_1 - \Delta_2) t} + \ket{2}\bra{1} \e^{i(\Delta_1 - \Delta_2) t}\right) 
\end{multline}
and for $j=2$ to
\begin{multline}
H_\mt{I3}(t) = \frac{\wh{p}^2}{2m} - \hbar\Delta_2 \ket{2}\bra{2}+  \frac{\hbar\Omega_2(\wh{x})}{2} \left(\ket{1}\bra{2}  + \ket{2}\bra{1} \right) \\+ \frac{\hbar\Omega_1(\wh{x})}{2} \left(\ket{1}\bra{2} \e^{i(\Delta_1 - \Delta_2) t} + \ket{2}\bra{1} \e^{-i(\Delta_1 - \Delta_2) t}\right) .
\end{multline}
In these interaction pictures the Hamiltonian is either constant or periodic with period $2\pi/(\Delta_1 - \Delta_2)$. For equal detuning, $\Delta_1 = \Delta_2$, these pictures would be time-independent and the standard result for $P_{12}$ would arise.


\section{Semiclassical solution of the Schr\"odinger equation}

For fast enough particles, i.e.\ for kinetic energies $E = mv^2/2 = \hbar^2 k^2/2m$ much larger than $\hbar\Omega$ and $\hbar\Delta_j$, the center-of-mass motion of the atom can be treated classically and independently of the internal dynamics. In that case the two-component wavefunction $\ket{\psi_\mt{I1}(t)}$ which accounts for the internal dynamics in the interaction picture I1 is a solution of the internal Schr\"odinger equation
\begin{equation} \label{eq:SE_internal}
 i\hbar \frac{d}{dt} \ket{\psi_\mt{I1}(t)} = H^\mt{scl}_\mt{I1}(t)\ket{\psi_\mt{I1}(t)},
\end{equation}
where
\begin{equation} \label{eq:Hscl}
H^\mt{scl}_\mt{I1} = \sum_{j=1,2} \frac{\hbar \Omega(x_0+vt)}{2} \bigl( \ket{1}\bra{2} \e^{i\Delta_j t} + \ket{2}\bra{1}\e^{-i \Delta_j t})
\end{equation}
and $x_0 <0$ is the position of the atom at time $t=0$.
In the following, we consider the internal dynamics for a given kinetic energy $E$, i.e.\ for a single atom whose center-of-mass follows the classical trajectory $x(t) = x_0 + vt$. We denote by $t_0$ the time of the first interaction with the leftmost field, $t_0 = -x_0/v$. This treatment neglects initial uncertainties in position and momentum and is well justified for atoms in a trap and a pulsed experiment. The case of a bunch of atoms with initial position and momentum distribution crossing to separated fields leads to a distribution of entrance times $t_0$ and it is considered later in Sec.~\ref{sec:packet}.

For an initial internal state $\ket{\psi_\mt{I1}(t_0)} = \ket{1}$ the solution of Eq.~(\ref{eq:SE_internal}) is given in terms of the evolution operator $U_\mt{I1}(t,t_0)$,
\begin{equation}
\ket{\psi_\mt{I1}(t)} = U_\mt{I1}(t,t_0) \ket{\psi_\mt{I1}(t_0)},
\end{equation}
where $U_\mt{I1}(t,t_0)$ fullfills
\begin{eqnarray} \label{eq:Udgl}
i\hbar \frac{d}{dt} U_\mt{I1}(t,t_0) &=& H^\mt{scl}_\mt{I1}(t) U_\mt{I1}(t,t_0),\nonumber\\ U_\mt{I1}(t_0,t_0) &=& U(t_0,t_0) = \wh{1}.
\end{eqnarray}
To obtain analytical results, we first consider the case of mesa mode functions for the two fields, $\Omega_1(x) = \Omega$ for $0\leq x \leq l$ and zero elsewhere and $\Omega_2(x) = \Omega$ for $l+L\leq x \leq 2l+L$ and zero elsewhere. In Sec.~\ref{sec:gaussmode} we will show the stability of our results with respect to more realistic field modes.
In the field free regions the Hamiltonian in this interaction picture is zero and thus the evolution operator is unity. Within the $j$th field, the solution of Eq.~(\ref{eq:Udgl}) is given by
\begin{widetext}
\begin{equation} \label{eq:UI1}
U_\mt{I1}^{(j)}(t,t_0) = \begin{pmatrix}
\e^{i\Delta_j(t-t_0)/2} \left\{\cos\left[ \frac{\Omega'_j(t-t_0)}{2} \right] - \frac{i\Delta_j}{\Omega'_j} \sin\left[ \frac{\Omega'_j(t-t_0)}{2} \right] \right\} & 
-\frac{i\Omega}{\Omega'_j} \e^{i\Delta_j (t+t_0)/2} \sin\left[ \frac{\Omega'_j(t-t_0)}{2} \right] \\ & \\
-\frac{i\Omega}{\Omega'_j} \e^{-i\Delta_j (t+t_0)/2} \sin\left[ \frac{\Omega'_j(t-t_0)}{2} \right] & \e^{-i\Delta_j(t-t_0)/2} \left\{\cos\left[ \frac{\Omega'_j(t-t_0)}{2} \right] + \frac{i\Delta_j}{\Omega'_j} \sin\left[ \frac{\Omega'_j(t-t_0)}{2} \right] \right\}
\end{pmatrix},
\end{equation}
\end{widetext}
where the effective Rabi frequencies $\Omega'_j = \left(\Omega^2 + \Delta_j^2\right)^{1/2}$, $j=1,2$, have been defined and $\ket{1} \equiv {1 \choose 0}$, $\ket{2} \equiv {0 \choose 1}$. 

Now assume that at time $t=t_0$ the ground state atom interacts with the first field for a time $\tau = l/v$, evolves freely a time $T = L/v$ and finally interacts with the second field for another time $\tau$. Thus, the final internal state is
\begin{multline} \label{eq:final_internal}
 \ket{\psi_\mt{I1}(t_0+2\tau+T)} = U_\mt{I1}^{(2)}(t_0+2\tau+T,t_0+\tau+T)\\ \times U_\mt{I1}^{(1)}(t_0+\tau,t_0) \ket{\psi(t_0)}.
\end{multline}
This yields for the probability of a transmitted excited state
\begin{eqnarray} \label{eq:P12_general}
P_{12}(\Delta_1,\Delta_2) &=& |\braket{2}{\psi_\mt{I1}(t_0+2\tau+T)}|^2 \\ 
&=& \Biggl| \e^{i(\Delta_1 - \Delta_2)(t_0+\tau)/2} \e^{-i\Delta_2 T/2} \sin\left(\frac{\Omega'_2 \tau}{2}\right) \nonumber\\
&&\times\left[\cos\left(\frac{\Omega'_1 \tau}{2}\right) - \frac{i \Delta_1}{\Omega'_1} \sin\left(\frac{\Omega'_1 \tau}{2}\right) \right] \frac{\Omega}{\Omega'_2} \nonumber\\
&& +~  \e^{-i(\Delta_1 - \Delta_2)(t_0+\tau)/2} \e^{i\Delta_2 T/2} \sin\left(\frac{\Omega'_1 \tau}{2}\right) \nonumber\\
&&\times \left[\cos\left(\frac{\Omega'_2 \tau}{2}\right) + \frac{i \Delta_2}{\Omega'_2} \sin\left(\frac{\Omega'_2 \tau}{2}\right) \right] \frac{\Omega}{\Omega'_1} \Biggr|^2.\nonumber\\&&
\end{eqnarray}
Note that $P_{12}(\Delta_1,\Delta_2)$ is periodic in the initial time $t_0$ with period $T = 2\pi/(\Delta_1-\Delta_2)$.

As a check we consider the limiting case of equal detuning, $\Delta_1 = \Delta_2$, leading to
\begin{multline}
P_{12}(\Delta,\Delta) = \frac{4\Omega^2}{\Omega'^2} \sin^2\left(\frac{\Omega' \tau}{2} \right) \left[\cos\left(\frac{\Omega' \tau}{2} \right) \cos\left(\frac{\Delta T}{2}  \right) \right.\\
\left. - \frac{\Delta}{\Omega'} \sin\left(\frac{\Omega' \tau}{2} \right) \sin\left(\frac{\Delta T}{2}  \right) \right]^2,
\end{multline}
which is independent of $t_0$ and coincides with the well-known result obtained by Ramsey \cite{Ramsey}. 

\begin{figure}
\begin{center}
\epsfxsize=8cm  \epsfbox{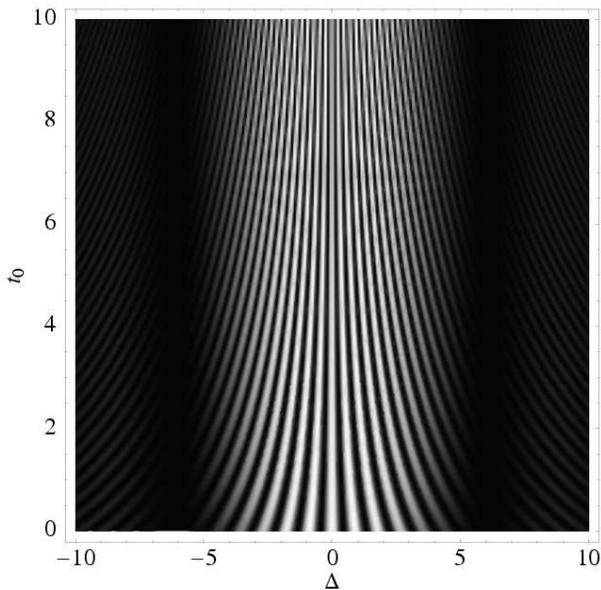}
\end{center}
\caption{Contour plot of $P_{12}(-\Delta,\Delta)$ as a function of $\Delta$ and $t_0$ for $\tau=1$, $T=5$. White color corresponds to a value of $1$ whereas black
corresponds to $0$. With increasing $t_0$, the interference pattern with respect to $\Delta$ becomes narrower. The zeros of $P_{12}(-\Delta,\Delta)$ at $\Delta \approx \pm 6$ are independent of $t_0$ and they are given by $\sin(\Omega' \tau/2) = 0$ according to Eq.~(\ref{eq:P12_minpl}). For all plots we use dimensionless units with $\hbar = m = 1$.}
\label{fig:P12contour}
\end{figure}

\begin{figure}
\centering
\epsfxsize=8.8cm  \epsfbox{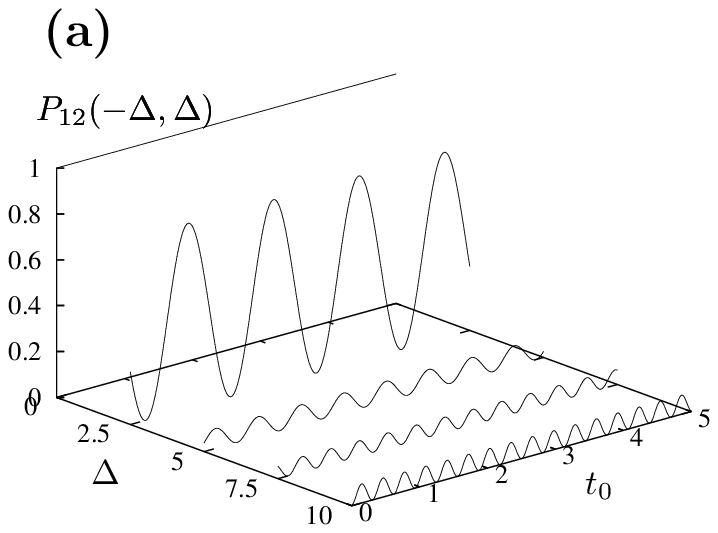}
\epsfxsize=8.8cm  \epsfbox{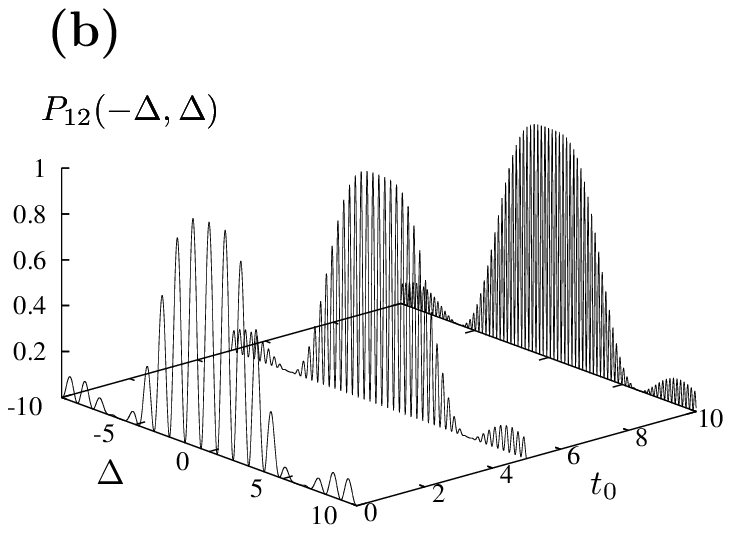}
\caption{(a) $P_{12}(-\Delta,\Delta)$ as a function of $t_0$ for $\Delta = \{0,2.5,5,7.5,10\}$. $P_{12}$ is periodic in $t_0$ with a period that depends on $\Delta$. We emphasize the fact that $P_{12}(0,0)=1$ independent of $t_0$ (straight line), indicating that the central peak remains stable irrespective of the entrance time. (b) $P_{12}(-\Delta,\Delta)$ as a function of $\Delta$ for $t_0 = \{0,5,10\}$, showing the narrowing of the standard Ramsey pattern. Parameter values are the same as in Fig.~\ref{fig:P12contour}.}
\label{fig:P12cuts}
\end{figure}

Another interesting case is that of equal modulus of detuning but opposite sign in both fields,
\begin{multline} \label{eq:P12_minpl}
P_{12}(-\Delta,\Delta) = \frac{4\Omega^2}{\Omega'^2} \sin^2\left(\frac{\Omega' \tau}{2} \right) \cos^2\Bigl[\Delta(t_0+\tau+T/2) \Bigr] \\ \times \left[ \cos^2\left(\frac{\Omega' \tau}{2} \right) + \frac{\Delta^2}{\Omega'^2} \sin^2\left(\frac{\Omega' \tau}{2} \right)\right],
\end{multline}
which maximizes the effect of different field frequencies.
We plot $P_{12}(-\Delta,\Delta)$ in Fig.~\ref{fig:P12contour} as a function of $\Delta$ and $t_0$. Remarkably, the interference pattern with respect to $\Delta$ becomes narrower if the initial entrance time $t_0$ is increased. This may appear astonishing at first sight since we expect periodicity in $t_0$. As it is shown in Fig.~\ref{fig:P12cuts}a, $P_{12}$ is indeed periodic in $t_0$ for fixed detuning, but with a detuning-dependent period $T=\pi/\Delta$, leading for increasing $t_0$ to a narrower Ramsey pattern as a function of $\Delta$, see Fig.~\ref{fig:P12cuts}b. An estimation for the width of the central fringe is obtained if one expands $P_{12}(-\Delta,\Delta)$ in a series around $\Delta=0$. Assuming a $\pi/2$-pulse for the fields, $\Omega = \pi/(2 \tau)$, this gives $P_{12}(-\Delta,\Delta) = 1- [T+2(t_0+\tau)]^2 \Delta^2/4 + \mathcal{O}(\Delta^3)$, such that the first zeros of the pattern are approximately given by
\begin{equation}
\Delta_0^\pm \simeq \pm \frac{2}{T+2(t_0+\tau)}.
\label{width}
\end{equation}
The central width is inversely proportional to the sum of the intermediate crossing time $T$ and the entrance time $t_0$. Of course, we could change the time $t_0$ by changing the origin of time. The important fact is that any such time shift would leave invariant the time {\it interval} $t_0$ between the instant in which the fields are in phase ($t=0$ in our time reference system so far) and the entrance of the atom.
Note the roles of $t_0$ and $T$ in Eq. (\ref{width}): First of all, $t_0$ is twice 
as efficient as $T$ to produce a desired width; moreover $t_0$ it is a time for free flight of ground state atoms, whereas $T$ is a free flight time for atoms with excited components, which are amenable of decay.  


An alternative viewpoint of the effect can be given by analyzing the phase mismatch between the two oscillating fields. This is better suited to understand the experiment with trapped ions and pulsed fields. Let us consider a more general version of the Hamiltonian (\ref{eq:Hscl}) where the phases of the fields at $t=0$ are given by $\phi_1$ and $\phi_2$, respectively,
\begin{multline}
H^\mt{scl}(t) = \hbar\omega_{21} \ket{2}\bra{2}\\ + \sum_{j=1,2} \frac{\hbar \Omega(x_0+vt)}{2} \bigl( \ket{1}\bra{2} \e^{i\omega_j t + i \phi_j} + \ket{2}\bra{1}\e^{-i \omega_j t -i \phi_j}).
\end{multline}
An equivalent derivation as above, but for an entrance time $t = 0$, yields for this Hamiltonian
\begin{multline}
P_{12}(-\Delta,\Delta) = \frac{4\Omega^2}{\Omega'^2} \sin^2\left(\frac{\Omega' \tau}{2} \right) \cos^2\Bigl[\Delta(\tau+T/2) + \phi_2-\phi_1\Bigr] \\ \times \left[ \cos^2\left(\frac{\Omega' \tau}{2} \right) + \frac{\Delta^2}{\Omega'^2} \sin^2\left(\frac{\Omega' \tau}{2} \right)\right].
\end{multline}
This agrees with Eq.~(\ref{eq:P12_minpl}) if one sets 
\begin{equation} \label{eq:phase_condition}
(\phi_2-\phi_1)/\Delta = t_0.
\end{equation}
Thus, the effect of different entrance times $t_0$ is equivalent to a fixed entrance time $t = 0$ but with an initial phase difference $\phi_2 - \phi_1$ of the fields. In particular, this means that in order to obtain one of the curves shown in Fig.~\ref{fig:P12cuts}b one has to change for every value of $\Delta$ this initial phase difference to fullfill Eq.~(\ref{eq:phase_condition}). In practice, it might be difficult to hold $t_0$ stable due to phase fluctuations of the fields. However, this will not affect the central fringe as we will show in the Section~\ref{sec:packet}.

\section{Atomic wave packet}
\label{sec:packet}

Up to now, the monochromatic case with a fixed entrance time $t_0$ has been considered. This holds true for the pulsed experiment and if the micromotion of the atom within the trap can be neglected. However, if the incoming ensemble of moving atoms is described by an initial position and momentum distribution,
the entrance and crossing times for the individual atoms will be different. In this case one has to integrate $P_{12}$ over all possible classical trajectories, weighted by a phase space distribution $W(x,k)$. 
As before we assume free motion for the center-of-mass, unperturbed by the fields.
In the following, we restrict our analysis to the case $-\Delta_1 = \Delta_2 \equiv \Delta$ and we assume $W(x,k)$ to describe a minimum uncertainty packet when its center impinges the origin $x=0$ at time $t=t^c_0$. Thus, $W(x,k)$ is given by
\begin{multline} \label{eq:Wigner}
W(x,k) = \frac{1}{2\pi\, \delta x\,\delta k} \exp\left[-\frac{(k-k_c)^2}{2(\delta k)^2}\right] \\ \times \exp\left\{-\frac{[x-\hbar k (t-t^c_0)/m]^2}{2(\delta x)^2}\right\},
\end{multline}
where $\delta x$ and $\hbar(\delta k)$ are the uncertainties of position and momentum, connected by $(\delta x)(\delta k) = 1/2$ and $\hbar k_c$ is the mean momentum.

Now in Eq.~(\ref{eq:P12_general}) one
has to replace $\tau$ by $l/v = ml/\hbar k$, $T$  by $L/v = mL/\hbar k$ and 
$t_0$ by a varying entrance time $t_0^c + x/v = t_0^c + mx/\hbar k$, where $x$ and $k$ are distributed according to Eq.~(\ref{eq:Wigner}). Finally, one obtains
\begin{widetext}
\begin{eqnarray} \label{eq:P12int1}
 \bigl\langle P_{12}(-\Delta,\Delta) \bigr\rangle &=& \int_{-\infty}^\infty dx \int_{-\infty}^\infty dk~W(x,k)\,P_{12}(-\Delta,\Delta) \\ \label{eq:P12int2}
&=& \frac{2\Omega^2}{\Omega'^2} \int_{-\infty}^\infty dk ~ g(k) \sin^2\left(\frac{m\Omega' l}{2\hbar k}\right) \left\{1 + \exp\left(-2m^2\Delta^2 (\delta x)^2/\hbar^2/k^2\right) \cos\left[2\Delta \left(\frac{ml}{\hbar k} + \frac{mL}{2\hbar k}\right) + 2\Delta t_0^c \right]\right\} \nonumber\\&& \quad\quad\quad\quad
\times \left[ \cos^2 \left(\frac{m\Omega' l}{2\hbar k}\right) + \frac{\Delta^2}{\Omega'^2}\sin^2 \left(\frac{m\Omega' l}{2\hbar k}\right) \right],
\end{eqnarray}
\end{widetext}
where $g(k) = \int dx\,W(x,k)$.
If one translates for every fixed velocity $v$ the position width $\delta x$ into an uncertainty of the entrance times, $\delta t_0 = \delta x/v$, one sees from the exponential in the integrand of Eq.~(\ref{eq:P12int2}) that for $\delta t_0$ much larger than the period $t_\mt{p}=\pi/\Delta$ the dependence on $t_0 = -x_0/v$ and thus the effect of narrowing disappears, whereas for a sudden entrance, $\delta t_0 \ll t_\mt{p}$, the integrand coincides with Eq.~(\ref{eq:P12_minpl}).

\begin{figure}
\centering
\epsfxsize=10cm  \epsfbox{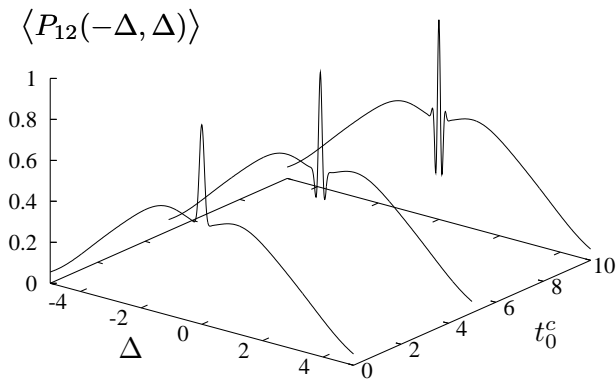}
\caption{$\langle P_{12}(-\Delta,\Delta) \rangle$ as a function of $\Delta$ for $t_0^c = \{0,5,10\}$, $l=1$, $L=5$. Parameters of the initial minimum uncertainty Gaussian distribution are: $k_c = 1$, $\delta k = 0.1$, $\delta x = 5$. The Rabi frequency has been chosen as a $\pi/2$-pulse for the mean velocity, $\Omega = \pi \hbar k_c/(2ml)$.}
\label{fig:P12_int}
\end{figure}

The $k$-integration has to be performed numerically and the result is shown in Fig.~\ref{fig:P12_int} as a function of $\Delta$ for fixed values of $t_0^c$. One clearly sees the Ramsey pedestal due to the averaging of the outer fringes, but the central fringe is not affected, it remains stable and becomes narrower for increasing $t_0^c$.

\section{Realistic mode function}
\label{sec:gaussmode}

\begin{figure}
\centering
\epsfxsize=10cm  \epsfbox{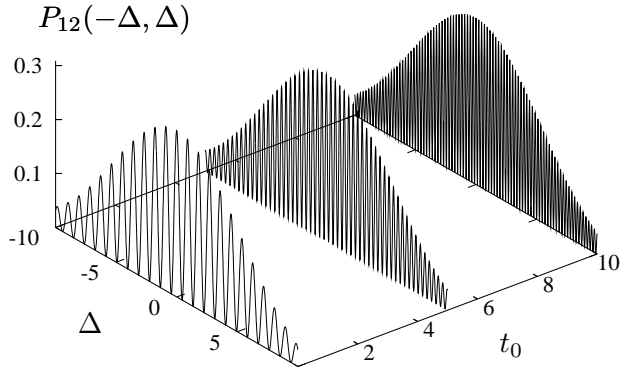}
\caption{$P_{12}(-\Delta,\Delta)$ as a function of $\Delta$ and $t_0 = \{0,5,10\}$ for realistic mode functions given in Eq.~(\ref{eq:realistic_mode}).
Parameters are: $\tau=1$, $T=5$, $\Omega = \pi/2$. The effect of narrowing is observable also for this mode functions.}
\label{fig:P12gaussmode}
\end{figure}

To investigate the stability of the narrowing effect with respect to the field mode function we withdraw the assumption of a mesa-shaped field within this section. As a more realistic example we consider a shape of the two field pulses which is very much reminiscent of a Gaussian but limited to a finite duration,
\begin{eqnarray} \label{eq:realistic_mode}
\Omega_1(t) &=& \displaystyle \Omega \sin^4\left(\frac{\pi (t-t_0)}{\tau}\right)
\chi_{[t_0,t_0 + \tau]}(t)\\ 
\Omega_2(t) &=& \displaystyle \Omega \sin^4\left(\frac{\pi (t-t_0-\tau-T)}{\tau}\right)\nonumber\\ &&\qquad\qquad\qquad\times \chi_{[t_0+\tau + T, t_0+2\tau+T]}(t),
\end{eqnarray}
where $\chi_{[t_1,t_2]}(t)=1$ for $t_1 \leq t \leq t_2$ and zero elsewhere.
These field shapes have also been used for STIRAP calculations \cite{FeShBe-AJP-1997}.
Again, the evolution operator between the two pulses is unity, whereas it has to be determined numerically by means of Eq.~(\ref{eq:Udgl}) in the field regions, where now the constant $\Omega$ has to be replaced by $\Omega_1(t)$ or $\Omega_2(t)$, respectively. The probability of excitation is calculated by means of Eqns.~(\ref{eq:final_internal}) and (\ref{eq:P12_general}) and the result is shown in Fig.~\ref{fig:P12gaussmode}. Irrespective of the sinusoidal mode, the effect of narrowing with respect to an increasing $t_0$ remains stable. Since the central peak is located at $\Delta=0$ for all values of $t_0$, a phase-space integration as in Section~\ref{sec:packet} will not change the central fringes and the result would be qualitatively similar as it has been shown in Fig.~\ref{fig:P12_int} for the case of mesa mode functions.

\section{Discussion}

In this paper we have studied the interference fringes in a Ramsey interferometer, for the case that the separated fields have different detuning. 
The excitation probability $P_{12}(\Delta_1,\Delta_2)$, derived within a semiclassical picture neglecting quantum reflections at the fields, depends on the entrance time $t_0$ of atom at the first field, or, more precisely, on the phase difference of the two fields at this time. Our main result is that the width of the Ramsey fringes decreases for increasing $t_0$. Moreover, we have shown that this effect remains stable for atomic clouds with an initial momentum and position distribution and for sinusoidal field mode functions.
A similar effect is achieved in a time domain configuration of atoms at rest
by controlling the laser field phases. 

It would be interesting to investigate the Ramsey fringes for differently detuned fields beyond the semiclassical approximation, i.e.\ taking into account quantum reflections at the fields for very slow (ultracold) atoms.
This has been shown to yield interesting effects in the case of equal detuning \cite{SeiMu-preprint-2006}. However, the quantum treatment of the time-dependent problem leads to difficulties in the calculation of the matching conditions between regions of constant potential and time-periodic potential. In fact, one has to introduce energy sidebands and to match them individually, as it has been done in the case of a one-channel oscillating barrier to model quantum traversal times \cite{Buettiker}. The solution of the two-channel case with two separated oscillating fields in an open problem so far.

\section*{Acknowledgments}

The authors thank D.\ Guery-Odelin for helpful discussion. This work has been supported by Ministerio de Edu\-ca\-ci\'on y Ciencia (BFM2003-01003) and UPV-EHU (00039.310-15968/2004). D.S.\ acknowledges a fellowship within the Postdoc-Programme of the German Academic Exchange Service (DAAD).

\end{document}